\documentclass[a4paper]{PoS}   	
\usepackage{graphicx}				
\usepackage{float}

\usepackage{caption}
\usepackage{subcaption}
\usepackage{wrapfig}
\usepackage{amssymb}
\usepackage{fontenc}
\usepackage{times}
\usepackage{mathptmx}

\title{Multi-TeV Energy Resolution Studies with VERITAS}
\author{\speaker{Rita Wells} for the VERITAS Collaboration\thanks{veritas.sao.arizona.edu}\\
Iowa State University\\
E-mail: \email{rmwells@iastate.edu}}
\abstract{This work aims to investigate the systematic uncertainty in gamma-ray spectra arising from saturation effects from bright images reconstructed by VERITAS.  The goal of the work is to improve or validate the energy resolution used for deriving gamma-ray spectra in the sub-TeV to multi-TeV energy regime with VERITAS.  Saturation from multi-TeV gamma-ray events affects the image brightness used to reconstruct the energy of the primary gamma ray, and potentially biases the energy estimate.  We discuss a method for investigating the energy resolution bias in these multi-TeV gamma-ray events by looking at showers with large core distances that do not saturate the cameras.}
\FullConference{35th International Cosmic Ray Conference - ICRC2017\\
Bexco, Busan, South Korea\\
12-20 July, 2017}
\ShortTitle{Energy Resolution with VERITAS}

\begin{document}
\section{Introduction}
The VERITAS four telescope array is located at the Fred Lawrence Whipple Observatory (FLWO) in southern Arizona (31 40N, 110 57W,  1.3km a.s.l.).  VERITAS observes Cherenkov light initiated by gamma rays in the atmosphere.  It observes sources producing gamma rays in the energy range of 85\ GeV to 30\ TeV, with an energy resolution of approximately 20\%, reaching a sensitivity of 1\% Crab in 25\ h~\cite{Park}.  The energy reconstruction of a primary gamma ray incident on the atmosphere is done by interpolating over lookup tables constructed from simulated gamma rays over the parameters of the total amount of detected light in the camera image, here called the image size, the impact distance, measured as the distance between the telescope and the shower core, and the zenith angle of the observations~\cite{Daniel}.  In general, as the primary gamma-ray energy increases, so does the size in each camera at a given impact distance (see Figure~\ref{fig1}).  Each camera consists of 499 photomultiplier tubes (PMTs), whose analog pulses are converted to digital traces by flash analog to digital converters (FADCs)~\cite{FADCs}.  An array-wide read-out of the PMT traces is triggered by a coincident signal in at least two telescopes~\cite{Park}.  VERITAS uses 8 bit unipolar FADCs, which saturate at 256\ digital counts (dc).  In order to extend the dynamic range from 256 to 1500 dc, a pixel trace going over 250 dc throws a switch to connect the FADC to a delayed low-gain channel that lowers the gain of the PMT trace by a factor of 6~\cite{FADCs}.  However, even in the low-gain, pixel saturation from extremely bright, high energy showers, can become a significant source of bias to the energy reconstruction.

When viewing a Cherenkov shower from the ground, taking into account the physical profile of the shower can lead to better understanding of the energy reconstruction.  The intensity of light in the image remains roughly flat from the shower core out to about 150\ m, at which point the light intensity drops exponentially.  As the shower core moves further from the telescope, the elliptical images become more elongated along the line from the telescope to the shower core and the amount of light in each image pixel decreases~\cite{LG}.  The shower maximum occurs lower in the atmosphere for more energetic showers due to the higher number of particles produced before they go below the critical energy.  However, statistical fluctuations can occur in first interaction depth and thus also shower maximum~\cite{LG,Gaisser}.  This effect can also become a source of energy bias.

This work aims to characterize the systematic uncertainty in the energy reconstruction of multi-TeV gamma rays falling at large core distances from the array. These showers, due to the drop in intensity at the edges of the Cherenkov shower, are largely unaffected by pixel saturation from extremely bright showers and have minimal light in low-gain channels; therefore, this technique circumvents the saturation effects, and thus provides an alternative analysis method for estimating systematic uncertainty.

\section{Investigating large core distance events}
\subsection{Methods}
To investigate the issue of image saturation, Monte Carlo simulations were run with Corsika and GrISUdet, both with and without the low-gain channels active.  Corsika simulates the Cherenkov shower; the simulated shower is then fed to GrISUdet~\cite{Duke} which simulates the VERITAS detector.  By turning off the low gain switch in half of the simulations, all pixels in those simulations that would have gone into low gain were allowed to saturate at a size of 256\ dc.  This allowed an investigation of where low gain began to affect the size of the image in various TeV images.  

The simulations ran from 30GeV to 30TeV, and additional mono-energetic simulations were generated at 800GeV, 1TeV, 3TeV, 5TeV, and 10TeV.  The showers were simulated out to core distances of 750m,  at a source position offset 0.5 degrees from the center of the camera, consistent with typical VERITAS observations.  The standard analysis package of VERITAS as described in~\cite{VEGAS} was used to reconstruct the simulated events.

\begin{figure}[H]
\centering
 	\includegraphics[width=0.8\textwidth]{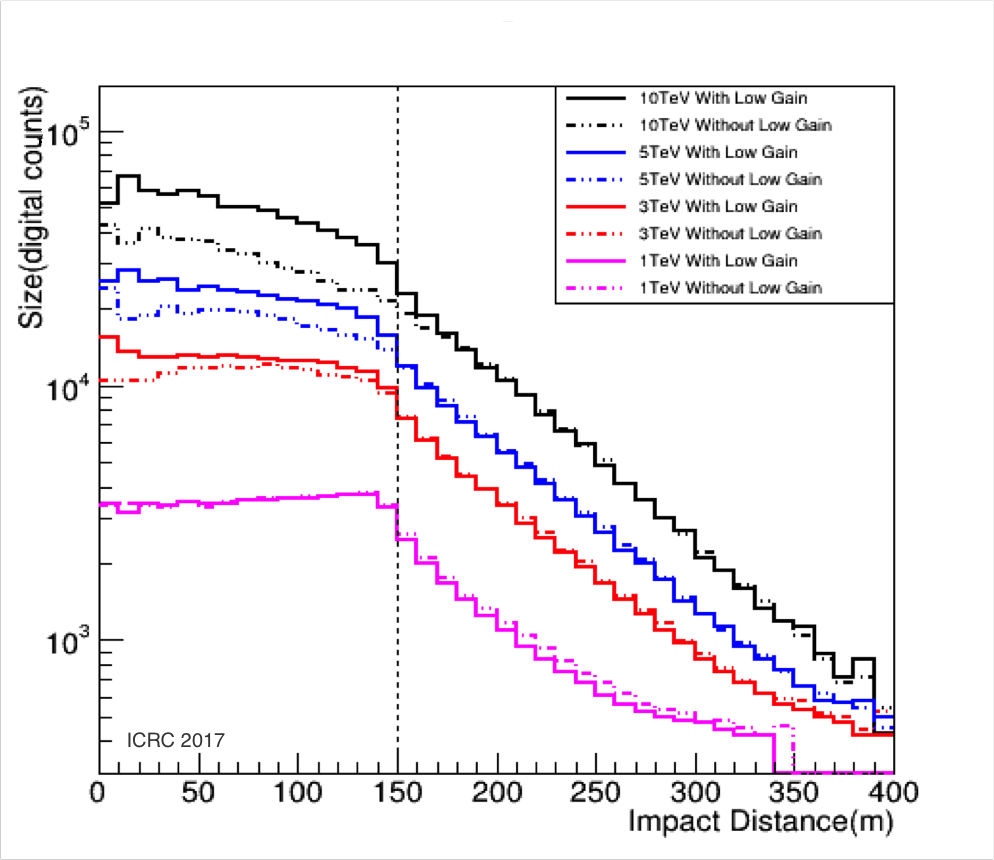}
  	\caption{\label{sm} Image size in a telescope vs. distance of core from telescope at 1, 3, 5, and 10\ TeV.  Solid lines are standard simulations, while dotted lines have low gain mode disabled and allow all image pixels to saturate at 256\ dc.}
  	\label{fig1}
\end{figure}

As seen in Figure~\ref{fig1}, plotting size vs.\ impact distance between each telescope and the shower core for a number of energies clearly shows that the low gain pixels have little affect on the size of the images below a size of $10^4$, corresponding to energies of a few TeV.  Even at energies up to 10TeV, once the telescope is beyond 150 meters from the shower core, saturation drops dramatically.

\begin{figure}[H]
 \centering
 \includegraphics[width=0.45\textwidth]{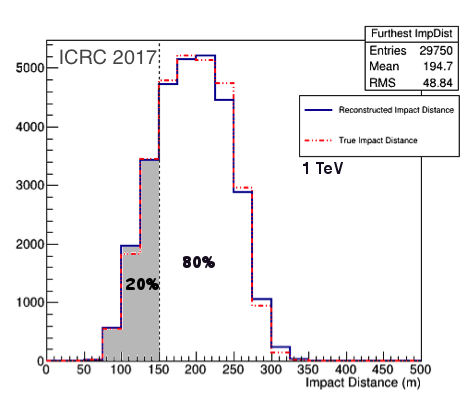}
 \includegraphics[width=0.45\textwidth]{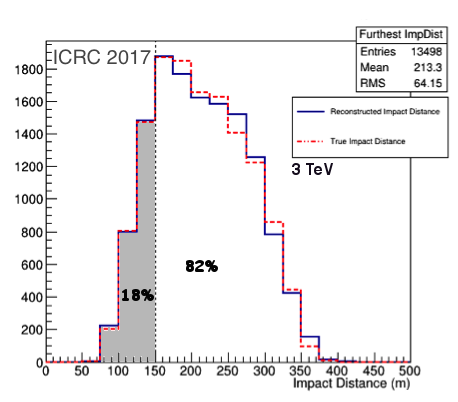}
 \includegraphics[width=0.45\textwidth]{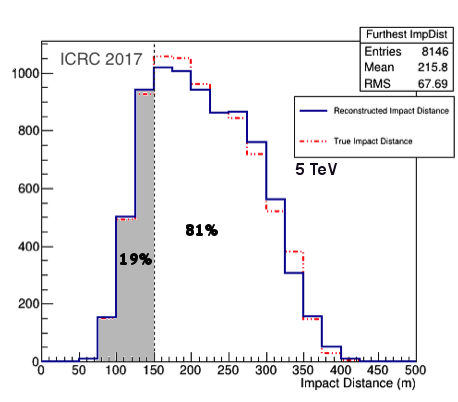}
 \includegraphics[width=0.45\textwidth]{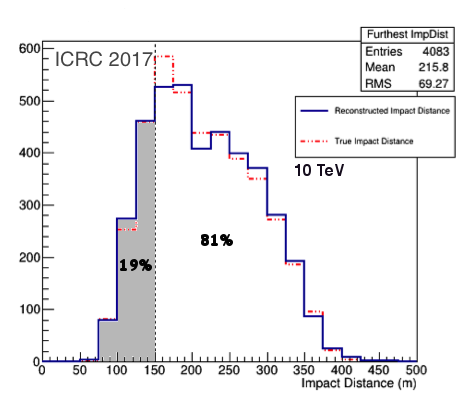}
 \caption{\label{sm} Plots showing the impact distance of the furthest telescope in the event at 1, 3, 5, and 10TeV.  Up to 10TeV, the efficiency of requiring an impact distance greater than 150m is at least 80\%.}
 \label{fig2}
\end{figure}

Figure~\ref{fig2} displays data as to whether there were enough images at multi-TeV energies that fell far enough from the telescopes to be unsaturated.  By plotting the impact distance of the furthest telescope from the shower core for every event, it is clear that about 80\% of multi-TeV events are recovered by selecting only distant events at high energies.

Nearly 100\% of events are kept at sub-TeV energies.  For all energies, the impact distance reconstruction is excellent with a resolution of ~10\% on the core reconstruction, regardless of saturation, out to about 300-350m.

\subsection{New analysis cuts}
As Figure~\ref{fig1} shows, the size of the image can be used as an initial cut on whether or not the image will have saturation.  Therefore, during the energy reconstruction, an extra quality cut is added to each telescope before energy reconstruction based on size and impact distance.  For any telescope with a size greater than $10^4$dc, only telescopes that are more than 150m from the shower core are used in the energy reconstruction.  The energy calculated for the remaining telescopes that pass the quality cuts in each event are then averaged together to obtain the reconstructed gamma-ray energy.

This extra cut leaves low energy showers largely unaffected, while multi-TeV showers lose about 20\% of events.  The remaining events that lose some telescopes in the energy reconstruction lose only highly saturated telescopes, so that the resulting average is unbiased by saturation effects.

The saturated pixels have no significant effect on the core reconstruction and other quality cuts.  Therefore, every telescope that passes the initial telescope level quality cuts, as described in~\cite{Daniel}, can still be used to calculate core location and to reject background.  Only during energy reconstruction are the saturated telescopes removed from consideration.  

\subsection{Energy bias}
\begin{figure}[H]
\centering
 \includegraphics[width=0.8\textwidth]{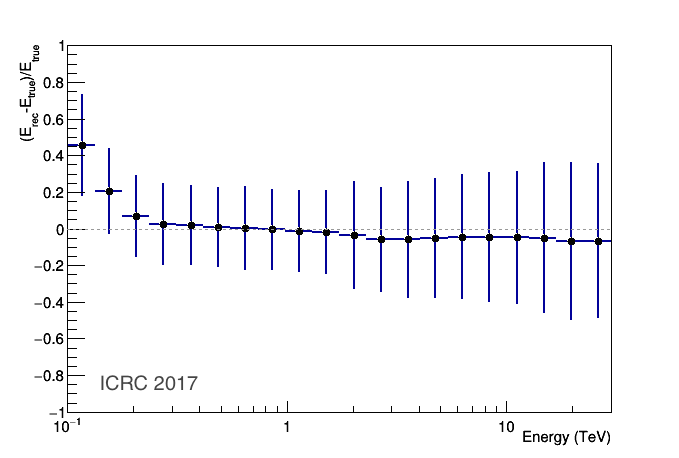}
 \caption{\label{sm} Plot of energy bias using the SAM analysis.}
 \label{fig3}
\end{figure}

At low energies, as expected, there is very little change in the energy resolution from the Saturation Avoidance Method (SAM) cuts.  At higher energies, rather than seeing a decrease in energy bias, the energy bias from avoiding saturated telescopes (see Figure~\ref{fig3}) is comparable to the energy bias from a standard analysis that includes saturated telescopes.  However, since the energy bias is calculated from tables comparing simulations to simulations, the results do not tell the whole story.

These results also indicate that other parameters, such as the shower maximum, have a much greater impact on the energy reconstruction bias than does saturation.

\begin{figure}[H]
\centering
 \includegraphics[width=0.75\textwidth]{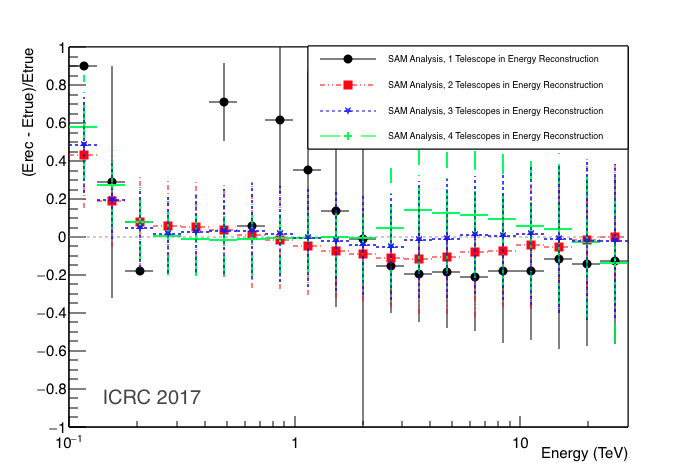}
 \caption{\label{sm} Plot of energy bias using the SAM analysis, broken down by the number of telescopes used in the energy reconstruction of the events. The 3-telescope set has almost no average energy bias at multi-TeV energies out to 30TeV.}
 \label{fig4}
\end{figure}

When the SAM analysis was broken down into 1,2,3, and 4 telescope subsets in Figure~\ref{fig4}, the 3 telescope subset showed a completely flat energy bias centered at zero bias for multi-TeV events, out to the limit of the VERITAS energy reconstruction.

The events in Figure~\ref{fig4} with only one telescope passing the additional cuts at the energy reconstruction show a large bias at both sub-TeV and TeV energies.  In order for an event to pass standard analysis cuts, at least two telescopes must pass the image quality cuts~\cite{VEGAS}.  Therefore, in any event with only one telescope passing the SAM cuts, there is at least one other telescope that was bright enough to be removed from the analysis.  Below multi-TeV energies, these images are the bright outliers that would bias the energy reconstruction towards higher energies.  At multi-TeV energies, only the dimmest events have only one telescope passing the SAM cuts.  The energy reconstruction of these dim outliers is biased towards lower energies.  Therefore, the events that reconstruct energy from only a single telescope in the SAM analysis are doubly biased across the VERITAS energy range.

\begin{figure}[H]
\centering
 \includegraphics[width=0.75\textwidth]{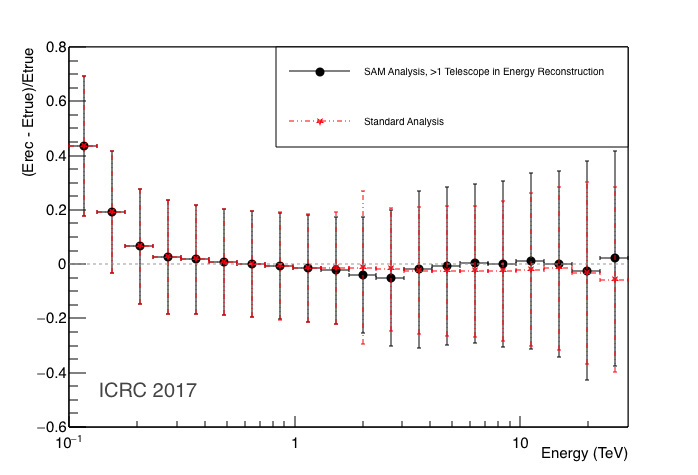}
 \caption{\label{sm} Plot of energy bias using the Saturation Avoidance Method analysis with the set of events containing only one telescope in the energy reconstruction removed.  The energy bias from the standard analysis is shown for comparison.}
 \label{fig5}
\end{figure}

Only 3.2\% percent of events have only one telescope used in the energy reconstruction, so removing the single telescope events from our energy reconstruction does not lead to a significant loss of statistics.  Removing the events with only one telescope passing the SAM analysis cuts does significantly improve the energy bias, as can be seen in Figure~\ref{fig5}.  Above about 3 TeV, the SAM analysis with at least two telescopes used in the energy reconstruction provides a slight improvement to the energy bias over the standard analysis.  The error bars show the energy resolution in each bin for Figure~\ref{fig3}, Figure~\ref{fig4}, and Figure~\ref{fig5}.  The energy resolution of the SAM analysis is consistent with the standard analysis; at the highest energies it is limited by statistics and thus slightly broadened over the standard analysis.

\section{Validation - The Crab Nebula}
The Crab Nebula has been observed by a number of IACTs and other gamma-ray instruments, including VERITAS, since its discovery.  It is one of the brightest sources in the gamma-ray sky, with no detected variability in the VERITAS energy range~\cite{Aliu}.  As such, it serves as an excellent standard source to validate the SAM analysis cuts outside of reliance on simulations.  While the standard analysis uses saturated images in the lookup tables used to reconstruct the energy of the primary gamma ray, the SAM analysis does not use those images.  For this reason, reconstructing real data with both the standard and the SAM analyses allows for a cross-check of the systematic effect of saturation on the energy reconstruction.

Investigation of the systematics on the VERITAS Crab spectrum are ongoing to obtain an accurate idea of the systematic bias deriving from saturated images in the standard VERITAS analysis.

\section{Summary}
This paper presents a method for avoiding saturated images in IACTs.  By removing extremely bright images near the shower core from the energy reconstruction, a sample of events unbiased by saturation effects can be reconstructed.  These events provide a method of probing the systematic uncertainty of any analysis methods aimed at reconstructing saturated events with a greater degree of accuracy.

\section{Acknowledgments}
This research is supported by grants from the U.S. Department of Energy Office of Science, the U.S. National Science Foundation and the Smithsonian Institution, and by NSERC in Canada. We acknowledge the excellent work of the technical support staff at the Fred Lawrence Whipple Observatory and at the collaborating institutions in the construction and operation of the instrument. The VERITAS Collaboration is grateful to Trevor Weekes for his seminal contributions and leadership in the field of VHE gamma-ray astrophysics, which made this study possible.


\begin{thebibliography}{99}


\bibitem{Park} N.\ Park. \emph{Performance of the VERITAS experiment}, in proceedings of \emph{34th ICRC}, \pos{PoS(ICRC2015)771} (2016).

\bibitem{Daniel} M.\ K.\ Daniel et al., \emph{The VERITAS standard data analysis}, in proceedings of \emph{30th ICRC}, \pos{PoS(ICRC2007)} (2007).

\bibitem{FADCs} P.\ F.\ Rebillot et al. \emph{The VERITAS Flash ADC Electronics System}, in proceedings of \emph{28th ICRC}, \pos{PoS(ICRC2003)} (2003).

\bibitem{LG} M.\ Lemoine-Goumard. \emph{3D-reconstruction of gamma-ray showers with a stereoscopic system.
Towards a Network of Atmospheric Cherenkov Detectors VII}, Apr 2005, Palaiseau, France.
pp.173--182c, 2005. <in2p3-00127358>

\bibitem{Gaisser} T.\ Gaisser. \emph{Cosmic Rays and Particle Physics}, Press Syndicate of the University of Cambridge, New York 1990.

\bibitem{Duke} C.\ Duke. [http://www.physics.utah.edu/gammaray/GrISU/].

\bibitem{VEGAS} P.\ Cogan. \emph{VEGAS, the VERITAS Gamma-ray Analysis Suite}, in proceedings of \emph{30th ICRC}, \pos{PoS(ICRC2007)} (2007).

\bibitem{Aliu} E.\ Aliu et al. \emph{A search for enhanced very high energy gamma-ray emission from the 2013 March Crab Nebula flare}, The Astrophysical Journal, 781, L11, 2013.

\end{thebibliography}
\end{document}